\documentclass[a4paper,12pt]{article}
 
\usepackage{amsfonts}
\usepackage{bm}
\usepackage{graphics}
\usepackage[dvips]{graphicx}

\newcommand{\beq}{\begin{equation}}
\newcommand{\eeq}{\end{equation}}
\newcommand{\beqs}{\begin{eqnarray}}
\newcommand{\eeqs}{\end{eqnarray}}

\newcommand{\mesm}{\widetilde{\mathcal{M}}}

\begin{document}

\begin{titlepage}
\hfill ULB-TH/07-25
\\
\vspace{25pt}

\begin{center}
{\LARGE \bf A Quiver of Many Runaways}

\end{center}

\vspace{15pt}

\begin{center}
{\large  Riccardo Argurio and Cyril Closset}\\

\vspace{25pt}
{\sl Physique Th\'eorique et Math\'ematique \\
and International Solvay Institutes \\
Universit\'e Libre de Bruxelles \\
CP 231, 1050 Bruxelles, Belgium}

\vspace{25pt}
{\tt rargurio, cyril.closset @ulb.ac.be}

\end{center}

\vspace{20pt}

\begin{center}
\textbf{Abstract}
\end{center}

We study the quantum corrections to the moduli space of the quiver gauge
theory corresponding to regular and fractional D3-branes at the 
$dP_1$ singularity. We find that besides the known runaway behavior 
at the lowest step of the duality cascade, there is a runaway direction along
a mesonic branch at every higher step of the cascade. Moreover, the algebra
of the chiral operators which obtain the large expectation values 
is such that we reproduce 
Altmann's first order deformation of the $dP_1$ cone.

\end{titlepage}
%%%%%%%%%%%%%%%%%%%%%%%%%%%%%%%%%%%%%%%%%%%%%%%%%%%%%%%%%%%%%%
%%%%%%%%%%%%%%%%%%%%%%%%%%%%%%%%%%%%%%%%%%%%%%%%%%%%%%%%%%%%%%
\renewcommand{\thefootnote}{\arabic{footnote}}
\setcounter{footnote}{0} \setcounter{page}{1}

%\newpage

\tableofcontents

\section{Introduction}
One of the most important advances in the study of the holographic
duality between gauge theories and string backgrounds was the generalization
of the AdS/CFT correspondence from D3-branes in flat space \cite{Maldacena}
to D3-branes
probing Calabi-Yau (CY) singularities \cite{Kachru,KW,MP}. 
When the latter singularities
are toric, a rigorous correspondence between the algebraic-geometric
properties of the singularity and the resulting quiver gauge theory
has been established over the years. In particular, it is very beautiful
to see how the complex equations characterizing the geometry arise
by solving for the classical moduli space of the $\mathcal{N}=1$ 
superconformal quiver gauge theory \cite{DGM,Berenstein}. 

In the case of the conifold singularity, it is known that there is a
complex deformation which leads to a smooth CY geometry, namely the
deformed conifold. The latter geometry also arises as a moduli space
of a quiver gauge theory \cite{KS}, 
in which however we have to depart from conformality
and introduce a non-trivial renormalization group (RG) flow. From the stringy
point of view, this is triggered by the presence of fractional branes.
It is argued that in the deep IR one ends up with a confining SYM
gauge theory, and the deformation parameter of the geometry is related 
to the gaugino condensate. In fact, as soon as some fractional branes are
included, the quantum moduli space of the quiver gauge theory separates
in several branches, each one representing a number of regular branes
probing the deformed conifold \cite{dks}. More specifically, the branches
associated to wandering 
regular branes are mesonic branches from the gauge theory
point of view, while the empty, smooth geometry is associated to 
the baryonic branch of the quiver gauge theory.

However, this behavior is not the typical one. In other geometries, fractional
branes trigger an RG flow which, after a cascade of Seiberg dualities, 
does not end in confining vacua, 
but rather in a theory which breaks supersymmetry with
a runaway behavior \cite{bhop,fhsu,bbc,is}, 
similarly to massless SQCD with $N_f<N_c$ \cite{ADS}. 
In the following, we will consider
the complex cone over the first del Pezzo surface, or in short $dP_1$, 
as the representative of such geometries. Its quiver gauge theory was
derived in \cite{Feng}. The $dP_1$ is known to have
an obstructed complex deformation \cite{Altmann1}, that is, a complex
deformation at first order which however has to vanish at second order
for consistency. 

In the present paper, we consider in detail the possible solutions
to the quantum F-term equations of the $dP_1$ quiver gauge theory. 
The classical
moduli space is consistently lifted everywhere, however we show that
the F-terms can be satisfied at infinity in field space on every mesonic
branch, signalling a runaway behavior. There are as many runaway
directions as there are steps in the duality cascade. Moreover, along the
runaway directions, the gauge invariants reproduce the equations of the
singularity deformed at first order. In other words, the regular D3-branes
are pushed to infinity, but as they run away, they are probing a geometry
corresponding exactly to the first order deformation of Altmann. Note that
the latter is not CY (and hence the background not supersymmetric) 
at quadratic order in the deformation, while of course  
the gauge theory breaks
supersymmetry because of the non-vanishing F-terms.
We hence observe a nice check of the gauge/string correspondence which
goes beyond the usual, protected, supersymmetric vacua but rather
extends to situations with only asymptotic supersymmetry.

The plan of the paper is the following. In Section 2 we consider the classical
moduli space of the $dP_1$ quiver gauge theory, paying attention to mesonic
and baryonic branches and their being decoupled. In Section 3 we derive
the quantum corrections, identifying all the runaway directions and
making the relation with the obstructed deformation of the geometry.
Some discussion is found in Section 4. In Appendix A, we apply
the same analysis as in the main text to study the various branches
of the moduli space of the conifold gauge theory, in order to 
``normalize'' our approach in a well-known example. In Appendix B, 
we make a similar analysis of the runaway mesonic branch for the quiver gauge
theory corresponding to supersymmetry breaking fractional branes at
the $dP_2$ singularity, for which we also derive the obstructed deformation.

\section{The classical moduli space of the $dP_1$ quiver gauge theory}
\begin{figure}[h]
\begin{center}
\includegraphics[height=6cm]{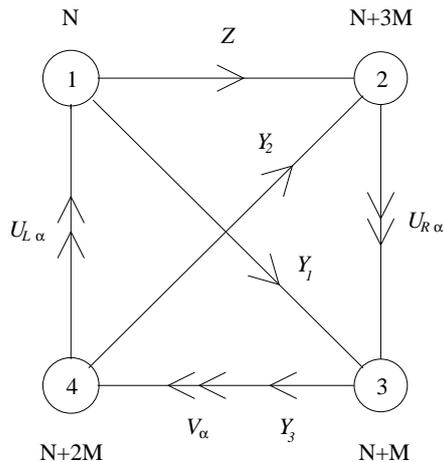}
\end{center} 
\caption{The $dP_1$ quiver for $N$ regular and $M$ fractional branes.}
\label{fig01}
\end{figure}

The quiver gauge theory corresponding to D3-branes probing a
$dP_1$ singularity has gauge group
$SU(N)\times SU(N+3M)\times SU(N+M) \times SU(N+2M)$ and matter fields
which can be read out from the diagram reproduced in figure \ref{fig01}.

The superpotential is
\beq
W_{tree}= h\mathrm{Tr}(\epsilon^{\alpha\beta} Y_3 U_{L\alpha}ZU_{R\beta}  
- \epsilon^{\alpha\beta} V_{\alpha}Y_2 U_{R\beta} + \epsilon^{\alpha\beta} 
V_{\alpha} U_{L\beta}Y_1), \label{wtree}
\eeq
where we choose to trace over the node 3 gauge group indices. 
Remark that we only have a diagonal $SU(2)$ flavor symmetry.
For later convenience, we already introduce variables which are suitable
for describing objects which are gauge invariant with respect to node 2,
the one with highest rank
\beq
\mathcal{M}_{\alpha}= ZU_{R\alpha} , \qquad \mathrm{and} \qquad  
\mathcal{N}_{\alpha}= Y_2 U_{R\alpha}. \label{gaugeinv}
\eeq
The classical F-term equations derived from extremizing (\ref{wtree})
are
\beqs
\epsilon^{\alpha\beta} U_{R\alpha}Y_3 U_{L\beta}&=& 0, \label{ft1}\\
\epsilon^{\alpha\beta}V_{\alpha}U_{L\beta} &=& 0, \label{ft2}\\
\epsilon^{\alpha\beta}U_{R\alpha} V_\beta &=& 0, \label{ft3}\\
\epsilon^{\alpha\beta}U_{L\alpha} ZU_{R\beta} &=& 0 , \label{ft4}\\
V_{\alpha} Y_2 & = & Y_3U_{L\alpha} Z , \label{ft5}\\
Y_1V_{\alpha} & = & ZU_{R\alpha}Y_3 ,\label{ft6} \\
Y_2 U_{R\alpha}  & = &  U_{L\alpha}Y_1.\label{ft7}
\eeqs

\subsection{The mesonic branch}
For generic $N$, we can build basic
``loops'' which consist of products of 3 or 4 bifundamentals such that
the resulting object has both indices in one gauge group \cite{bhop}. 
We have a total of 12 loops going through nodes 1-2-3-4, 6 loops going
through nodes 1-3-4 and 6 loops through nodes 2-3-4. It would thus seem
that if we base ourselves on nodes 1 or 2 we will see less loops and 
possibly a reduced moduli space. However this is not true because the F-terms
reduce the number of independent loops to 9, and eventually equate 
the eigenvalues of the loops based on different nodes. 
We briefly sketch below how this happens. See also Appendix A where the
same approach is applied in all details to the conifold gauge theory.

Let us for definiteness base ourselves on node 3. We immediately see that
the loop matrices will be distinguished by the number of $SU(2)$ indices
that they carry: one, two or three. Using the F-terms, we have respectively
\beq
Y_3 U_{L\alpha}Y_1 = Y_3 Y_2 U_{R\alpha}, 
\eeq
\beq
V_\alpha U_{L\beta}Y_1 =V_\alpha Y_2 U_{R\beta} = Y_3U_{L\alpha} Z
U_{R\beta} = Y_3U_{L(\alpha} ZU_{R\beta)},
\eeq
\beq
V_\alpha U_{L\beta} ZU_{R\gamma} = V_{(\alpha} U_{L\beta} ZU_{R\gamma)}.
\eeq
In particular, we see that all the $SU(2)$ indices are symmetrized
because of the first four F-term relations. We thus end up indeed
with 9 elementary loops, which we can name as follows, using
the gauge invariants of node 2 introduced in (\ref{gaugeinv})
\begin{eqnarray}\label{abc9}
a_1 = Y_3 \mathcal{N}_1,  \quad & b_1 = Y_3 U_{L 1}\mathcal{M}_1,\quad & c_1=
V_1U_{L 1}\mathcal{M}_1, \nonumber \\
a_2 = Y_3 \mathcal{N}_2, \quad & b_2 = Y_3 U_{L 1}\mathcal{M}_2,\quad & c_2=
V_1U_{L 1}\mathcal{M}_2,\nonumber \\
& b_3 = Y_3 U_{L 2}\mathcal{M}_2,\quad & c_3= V_2U_{L 1}\mathcal{M}_2, 
\nonumber \\
& & c_4= V_2U_{L 2}\mathcal{M}_2. \label{basicloops}
\end{eqnarray}
These 9 matrices commute, as one can easily check, 
so we can diagonalize them all. Moreover, they
are not independent. There are 20 quadratic 
relations between them, defining the 
complex cone over the first del Pezzo as a 3 dimensional affine  
variety in $\mathbb{C}^9$ \cite{bhop}:
\beq \label{algdescrdP1}
\begin{array}{cccc}
\quad a_1 b_2 = a_2 b_1 &
\quad a_1 b_3 = a_2 b_2 &
\quad b_2^2 = b_1 b_3 &
\quad b_2^2 = a_1 c_3 \\
\quad b_2^2 = a_2 c_2 &
\quad b_1^2 = a_1 c_1 &
\quad b_3^2 = a_2 c_4 &
\quad a_1 c_2 = b_1 b_2 \\
\quad a_1 c_4 = b_2 b_3 &
\quad a_2 c_1 = b_1 b_2 &
\quad a_2 c_3 = b_2 b_3 &
\quad b_1 c_2 = b_2 c_1 \\
\quad b_1 c_3 = b_2 c_2 &
\quad b_1 c_4 = b_2 c_3 &
\quad b_2 c_3 = b_3 c_2 &
\quad b_2 c_4 = b_3 c_3 \\
\quad b_2 c_2 = b_3 c_1 &
\quad c_1 c_4 = c_2 c_3 &
\quad c_2^2 = c_1 c_3 &
\quad c_3^2 = c_2 c_4 
\end{array}
\eeq
As complicated as they look, all the above relations can easily be
seen to arise just by considering that all quadratic objects with the
same $SU(2)$ indices must coincide and be symmetrized.

Exactly the same conclusion can be reached considering loops on any one
of the other three nodes. Note that because of their definitions (and because
of the above equations), all loops are eventually matrices of rank $N$
even when they are based on nodes of higher rank. Moreover, the most
generic situation is when all non vanishing eigenvalues are, say, 
in the upper-left corner.

In conclusion, 
the moduli space is a $N$-symmetric product of the CY affine variety.

We write now an explicit parametrization for the classical moduli space, 
which means solving for the D-terms and F-terms simultaneously. 
Our approach has been of course to first
solve the F-flatness conditions and then worry about the D-terms.

We adopt the solution of the F-terms given above by $9$ mutually 
commuting matrices (\ref{abc9}) at every node. Choose for instance the loops
\beq
a_1^{(3)} = Y_3Y_2U_{R1}, \qquad a_1^{(4)} = Y_2U_{R1}Y_3,
\eeq
based at node 3 and 4, respectively. Our loops obviously cannot have rank 
larger than $N$. Now use gauge freedom from gauge groups 3 and 4 to gauge fix 
\beq
(a_1^{(3)})^i_j = (a_1^{(3)})^i \delta^i_j, \quad 
(a_1^{(4)})^i_j = (a_1^{(4)})^i \delta^i_j, \qquad 
\mathrm{for}\quad i,j\leq N,
\eeq
zero otherwise. Now, $a_1^{(3)} Y_3 = Y_3 a_1^{(4)}$ implies that, for generic
vevs, $Y_3$ is diagonal 
in the upper-left $N\times N$ corner. One has (without summation):
\beq
(a_1^{(3)})^i Y^i_{3 j} = Y^i_{3 j} (a_1^{(4)})^j  \quad \Rightarrow \quad
Y^i_{3j} = 0 \quad \mathrm{if} \quad i\neq j \quad \mathrm{and} \quad
(a_1^{(3)})^i = (a_1^{(4)})^i
\eeq
for $i,j\leq N$, while for $p>N$,
\beq
(a_1^{(3)})^i Y^i_{3 p} = 0,   \qquad  Y^p_{3 i} (a_1^{(4)})^i  = 0.
\eeq
Note that the components $Y^p_{3 q}$ for $p,q>N$ remain undetermined.

By similar arguments, we can arrive at the conclusion that all sets 
of basic loops based on different nodes actually share the same
eigenvalues, and all elementary fields have a diagonal upper-left $N\times N$
part and an undetermined lower-right piece whose dimension is
$M\times 2M$ for $Y_3, V_\alpha$, $2M \times 3M$ for $Y_2$ and
$3M \times M$ for $U_{R\alpha}$.
To summarize, 
we have thus shown that all the bifundamental fields must have the form 
\beq \label{paramULD}
X =  \bordermatrix{&    &           \cr
                &X^D_{N\times N}& 0    \cr          
                &0    & \tilde{X} \cr      
                }.    
\eeq

We assume for the moment that $\tilde{X}=0$ for all the fields, 
so that all the vevs are diagonal. 
We list the additional constraints from the D-equations:
\begin{eqnarray}
|Z_i|^2 + |Y_{1i}|^2-|U_{L1i}|^2-|U_{L2i}|^2 &=&0 ,\label{d1}\\
|U_{R1i}|^2+|U_{R2i}|^2 -|Z_i|^2 -|Y_{2i}|^2 &=&0,\\
|V_{1i}|^2+|V_{2i}|^2 +|Y_{3i}|^2-|Y_{1i}|^2-|U_{R1i}|^2-|U_{R2i}|^2 &=&0,\\
|Y_{2i}|^2+|U_{L1i}|^2+|U_{L2i}|^2-|V_{1i}|^2-|V_{2i}|^2 -|Y_{3i}|^2 &=&0,
\label{d4}
\end{eqnarray}
where $i=1,..N$ runs over the upper-left diagonal blocks. The 
pattern of higgsing of the gauge group is
\beqs
G &=& SU(N)\times SU(N+3M)\times SU(N+M)\times SU(N+2M)  \nonumber \\
& \supset & SU(N)_\mathrm{diag} 
\times SU(3M)\times SU(M)\times SU(2M)  \nonumber \\
& \supset & U(1)^{N-1} \times SU(3M)\times SU(M)\times SU(2M),
\eeqs
where the non-abelian part is the $dP_1$ quiver for $N=0$ (i.e.~the
triangle quiver), while the $U(1)$'s 
are diagonal combinations of the Cartan subalgebras  of the four nodes' 
$SU(N)$ subgroups.

\subsection{The baryonic branches}
Let us now consider the special case $N=M$, which we take as
a case study of the more general situation $N=kM$. In this case we can
define baryonic gauge invariants for the second node with $SU(4M)$
gauge group, since it has effectively $N_f=4M$.\footnote{Actually, 
baryonic gauge invariants can generally be defined, for any node, when 
$N=k M$. However only when we have
$N_f=N_c$ for one node do the baryons become elementary effective fields 
at the quantum level.}

The mesonic gauge invariants of node two are $\mathcal{M}_\alpha$
and $\mathcal{N}_\alpha$ as defined in (\ref{gaugeinv}). They are respectively
pairs of $M\times 2M$ and $3M \times 2M$ matrices. We can thus
define a $4M \times 4M$ mesonic matrix as
\beq
{\mesm} \equiv \left( \begin{array}{cc} \mathcal{M}_1 & 
\mathcal{M}_2 \\
\mathcal{N}_1 & \mathcal{N}_2  \end{array}\right).
\eeq
More generally, we can define the same matrix also when $N\neq M$
and it will be $(2N+2M)\times (2N+2M)$.
In the mesonic branch considered above, it is clear that the matrices
$\mathcal{M}_\alpha$ and $\mathcal{N}_\alpha$ are non zero only in the
upper left rank $N$ part. Hence, $\mesm$ will be of maximal
rank $2N$ and $\det\mesm =0$. Actually, from the classical
F-terms (\ref{ft7}) we see that the matrices $\mathcal{N}_\alpha$ are
each of maximal rank $N$ (since there is a summation over $SU(N)$ indices
on the r.h.s.) so that, when $N=M$ we necessarily have 
$\det \mesm=0$ in any SUSY vacuum. This is going to play
an important role in the subsequent analysis.

In the $N=M$ case we can define two baryonic invariants, 
\begin{eqnarray}
\mathcal{B} &\propto &(Y_2)^{3M}Z^M \equiv \det \left(
\begin{array}{c} Z \\ Y_2 \end{array}\right),\\
\bar \mathcal{B} &\propto & (U_{R1}U_{R2})^{2M}\equiv \det
\left( \begin{array}{cc} U_{R1} & U_{R2}\end{array}\right)   ,
\end{eqnarray}
where both matrices entering the definitions are $4M \times 4M$.
Again, it is easy to see that on the mesonic branch $\mathcal{B},
\bar \mathcal{B} =0$ because of the non-maximal ranks of the matrices
involved.

We now ask whether it is possible to have regions or branches of the moduli
space where the baryonic invariants are turned on. We see that we can 
use the F-term (\ref{ft5}) in order to form gauge invariants involving
$\mathcal{B}$. We get the equations
\beq
V_\alpha \mathcal{B} = 0 = Y_3 U_{R\alpha}\mathcal{B}.
\eeq
This means that if $\mathcal{B}\neq 0$, then $V_\alpha= Y_3 U_{R\alpha}=0$
identically. It is easy to show that this in turn implies
that all basic loops (\ref{basicloops}) vanish. Similarly, 
from the F-terms (\ref{ft1}) and (\ref{ft3}) we obtain
\beq
V_\alpha \bar \mathcal{B} = 0 = Y_3 U_{R\alpha}\bar \mathcal{B},
\eeq
with the same conclusion of vanishing loops. Moreover, if $ \mathcal{B}\neq 0$
then $\bar \mathcal{B}$ has to vanish and vice-versa because
$\mathcal{B}\bar \mathcal{B} = \det \mesm=0$.

Thus we see that additionally to the mesonic branch, which consists of
$M$ symmetrized copies of the complex cone over $dP_1$, we have two
one-complex dimensional baryonic branches. All of the branches
of the moduli space meet at the origin. 

In order to see what is the left over gauge group on the baryonic branches, 
we have to solve for the D-terms.\footnote{Note that in this case the r.h.s. 
of the D-equations (\ref{d1})--(\ref{d4}), i.e.~the trace part, is non 
vanishing.}
Because the loops are all zero, we are
more constrained than on the mesonic branch and the elementary fields
will have VEVs proportional to the identity. More explicitely, when 
$ \mathcal{B}\neq 0$ we turn on only the $Z$ and $Y_2$ fields. 
It turns out that we have to take the left $M\times M$ part of $Z$
and the right $3M\times 3M$ part of $Y_2$ proportional to the identity,
with the same constant of proportionality. The gauge group is
broken according to the following pattern:
\beqs
G&=& SU(M)\times SU(4M)\times SU(2M)\times SU(3M)  \nonumber \\ 
 & \supset & SU(M)\times (SU(M) \times SU(3M)) \times SU(2M)\times SU(3M) 
\nonumber \\
& \supset & SU(M)_\mathrm{diag} \times SU(3M)_\mathrm{diag}\times SU(2M).
\eeqs  
Thus, we get the triangular quiver, and the matter content can be checked
to be the expected one by standard higgsing arguments. Note that on the
baryonic branch we do not have a $U(1)^{N-1}$ factor.

Similarly, on the $ \bar \mathcal{B}\neq 0$ branch we have that 
$U_{R1}$ and $U_{R2}$ have respectively their upper and lower
$2M\times 2M$ parts proportional to the identity because of the D-terms.
The gauge group is broken according to
\beqs
G &= & SU(M)\times SU(4M)\times SU(2M)\times SU(3M)  \nonumber \\
 & \supset & SU(M)\times (SU(2M) \times SU(2M)) \times SU(2M)\times SU(3M) 
\nonumber \\
& \supset & SU(M) \times SU(2M)_\mathrm{diag}\times SU(3M),
\eeqs
again obtaining the same theory, albeit embedded in a different 
way in the original gauge group.

Note that the fact that at any point of the various branches of the moduli
space we still have a non-trivial gauge theory, namely the triangular
quiver, means that each one of this points actually corresponds
to a moduli space of its own. In other words, every point of the moduli
space discussed here is itself a moduli space, which is the one discussed
in detail in \cite{is,abcc}.

\section{The quantum corrections to the $dP_1$ moduli space}
Here we wish to study how the classical picture is modified by quantum
corrections. The story in the $N=kM$ case is by now well-known \cite{bhop, 
fhsu,bbc}. 
The gauge theory is non conformal and is believed to undergo a non-trivial
RG flow which takes the form of a cascade of Seiberg dualities. The latter
effectively reduce the ranks at every node by $M$ at every step.\footnote{That 
this RG flow has to be the one described by a gravity dual
such as the ones in \cite{fhhw,hek} has 
been argued in \cite{ekk}.} 
At the last step, one usually goes to the (quantum)
baryonic branch and ends up with the triangular quiver, which is runaway
as we will rederive later (see \cite{is,abcc} for a discussion on how
one might stop this runaway behavior).

In the language of the previous section, the above result can be stated
by saying that the baryonic branch of the second node becomes runaway
because of quantum corrections coming from another node (the fourth).
Here we wish to address the question of what becomes of the mesonic
branch of the second node. Because the low energy gauge group is still
the triangular quiver, we also expect a runaway behaviour, but since
the embedding of the gauge group is different the runaway will be driven
by different quantum effects. Also, from the dual stringy perspective, 
on the mesonic branch we have regular D3-branes around and the question
is whether they will feel a potential, or what space they will seem to
be probing.

In the following, we start by considering the case $N=M$ which hopefully
captures most of the physics we want to discuss. We will turn later to
the more general case $N\neq M$.

\subsection{Runaway on the baryonic branch}
The effective superpotential for $N=M$ is
\beq 
W= h\mathrm{Tr}(\epsilon^{\alpha\beta} Y_3 U_{L\alpha}\mathcal{M}_{\beta}  
- \epsilon^{\alpha\beta} V_{\alpha}\mathcal{N}_{\beta} + 
\epsilon^{\alpha\beta} V_{\alpha} U_{L\beta}Y_1)+L(\det{\mesm} -
\mathcal{B} \bar \mathcal{B}- \Lambda_2^{8M}), \label{wcons}
\eeq
where $L$ is a superfield Lagrange multiplier.

The F-terms are the following
\beqs
\det{\mesm} - \mathcal{B} \bar \mathcal{B}& = & \Lambda_2^{8M} , \label{ef1}\\
L \mathcal{B} = & 0 & = L \bar \mathcal{B},\label{ef2} \\
\epsilon^{\alpha\beta}  Y_3 U_{L\alpha}  & = & - L
{\partial \det{\mesm} \over \partial \mathcal{M}_{\beta}}, \label{ef3}\\
\epsilon^{\alpha\beta} V_{\alpha} & = & L 
{\partial \det{\mesm} \over \partial \mathcal{N}_{\beta}}, \label{ef4}\\
\epsilon^{\alpha\beta} U_{L\alpha} \mathcal{M}_{\beta} &= & 0, \label{ef5}\\
\epsilon^{\alpha\beta} V_\alpha U_{L\beta}  &= & 0, \label{ef5bis}\\
\mathcal{M}_{\alpha} Y_3 &= & Y_1 V_{\alpha}, \label{ef6}\\
\mathcal{N}_{\alpha} &= & U_{L\alpha} Y_1.\label{ef7}
\eeqs
As in the classical case, if we want to satisfy (\ref{ef7}), then
the matrix $\mesm$ is not of maximal rank and $\det{\mesm}=0$. We are
then automatically on the baryonic branch: the constraint (\ref{ef1})
forces the baryons $\mathcal{B}, \bar \mathcal{B}$ to have non zero
VEVs (at the quantum level they must be both non vanishing), which
in turn implies $L=0$ from (\ref{ef2}). Then, (\ref{ef3}) 
and (\ref{ef4}) imply that $V_{\alpha}$
and $ Y_3 U_{L\alpha}$ are zero, which eventually means that all the
loop variables are zero. We are definitely on the baryonic branch which, 
as far as the dynamics of node two is concerned, is still supersymmetric.
At this stage, note that the mesonic branch has no chance of appearing
because non-vanishing loops would need $L\neq 0$ which would mean vanishing
baryons and $\det{\mesm}=\Lambda_2^{8M}$, contradicting one of the F-terms.

So, we see that quantum effects at node two lift the mesonic branch but not
the baryonic one (which is the smooth merger of the two classical baryonic
branches).

\begin{figure}[h]
\begin{center}
\includegraphics[height=6cm]{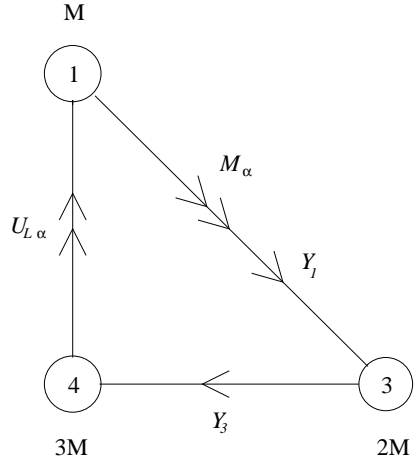}
\end{center} 
\caption{The last step of the cascade of the 
$dP_1$ quiver for $M$ fractional branes.}
\label{fig02}
\end{figure}
For the sake of completeness, we reproduce here the well-known result that
the baryonic branch is also eventually lifted by quantum corrections. 
Indeed, on the baryonic branch we are left with a triangular quiver
with gauge group $SU(M)\times SU(2M)\times SU(3M)$ and matter represented
by $U_{L \alpha} $, $Y_1$, $Y_3$ and $\mathcal{M}_{\alpha}$
($V_{\alpha}$ and $\mathcal{N}_{\alpha}$ have been integrated out
because they appear quadratically in (\ref{wcons})). The quiver is
represented in Figure \ref{fig02}.
The matter content is such that node four with gauge group $SU(3M)$
has $N_f=2M < N_c$ flavors. Hence, an ADS-like effective superpotential will
be generated for its mesons $X_\alpha =Y_3 U_{L\alpha}$
\beq
W_\mathrm{eff}= h \mathrm{Tr}\epsilon^{\alpha\beta} X_\alpha\mathcal{M}_{\beta}
+ M \left( {\Lambda_4^{7M} \over \det \tilde X}\right)^{1\over M},
\eeq
where we have defined the $2M\times 2M$ matrix $\tilde X \equiv ( X_1 \,
X_2 )$ and $\Lambda_4$ is the dynamical scale of node four.
It is clear that the F-terms will set $X_\alpha$ to zero while sending
$\mathcal{M}_{\alpha}$ to infinity. This is the runaway direction
at the last step of the cascade. In the following, we want to see if there
are other, disconnected runaway directions corresponding to the other branches
of the classical moduli space.

\subsection{Runaway on the mesonic branch}
For simplicity, we consider here solutions to the F-terms 
(\ref{ef1})--(\ref{ef7}) in the special case $N=M=1$. We force being on the
mesonic branch by requiring that $L\neq0$ so that $\mathcal{B}, 
\bar \mathcal{B} = 0$ and 
\beq
\det\mesm = \Lambda^8.
\eeq
We immediately see that, if we are to find a solution to the F-terms,
it will be runaway, because the above condition conflicts with the
rank condition following from (\ref{ef7}). Hence, both equations
will be satisfied only if some elements of $\mesm$ go to infinity
as others go to zero. The non-trivial task is to find a scaling
for all the fields appearing above such that all F-terms go to zero
while loop variables remain non zero. The Lagrange multiplier $L$ 
should also be large enough in order for this
branch to be really disconnected from the baryonic one, as it is the
case classically.

All fields will thus have a non zero VEV assigned to their upper-left
component, which is the one entering in the loop variables. Additionally, 
at least the fields $\mathcal{N}_\alpha$ will have to have some non-zero
component in the lower-right part. We will see that as a consequence also
$V_\alpha$ will need to have such a component.
We thus take
\beq
\mathcal{M}_\alpha = 
\left(\begin{array}{cc} m_\alpha & 0 \end{array}\right) , \qquad
\mathcal{N}_\alpha = 
\left(\begin{array}{cc} n_\alpha & 0 \\ 0 & \epsilon_\alpha \\
0 & \delta_\alpha \end{array}\right) , \label{ans1}
\eeq
\beq
Y_3 = \left(\begin{array}{ccc} y_3 & 0 & 0 \\ 0 & 0 & 0 \end{array}\right),
\qquad V_\alpha = 
\left(\begin{array}{ccc} v_\alpha & 0 & 0 \\ 0 & w_\alpha & 
x_\alpha \end{array}\right),\label{ans2}
\eeq
\beq
Y_1 = \left(\begin{array}{cc} y_1 & 0 \end{array}\right) , \qquad
U_{L\alpha}= \left(\begin{array}{c} u_\alpha \\ 0 \\ 0 \end{array}\right).
\label{ans3}
\eeq
It is convenient to rewrite the effective superpotential (\ref{wcons})
in terms of the above ansatz
\beq
W= y_3 u_\alpha m^\alpha -v_\alpha n^\alpha - w_\alpha \epsilon^\alpha
-x_\alpha \delta^\alpha + v_\alpha u^\alpha y_1
-L (m_\alpha n^\alpha \epsilon_\beta \delta^\beta +\Lambda^8),
\eeq
where all fields appearing are no longer matrices, and e.g. $m^\alpha\equiv
\epsilon^{\alpha\beta}m_\beta$.

The F-terms simply read
\beqs
u_\alpha m^\alpha &=& 0, \label{sf1}\\
v_\alpha u^\alpha &= & 0, \label{sf2}\\
y_3 m_\alpha &= & v_\alpha y_1 , \label{sf3}\\
n_\alpha & = &  u_\alpha y_1, \label{sf4}\\
\epsilon_\alpha &=& 0, \label{sf5}\\
\delta_\alpha &= & 0, \label{sf6}\\
y_3 u_\alpha &= & -L n_\alpha  \epsilon_\beta \delta^\beta, \label{sf7}\\
v_\alpha &= & -L m_\alpha  \epsilon_\beta \delta^\beta, \label{sf8}\\
w_\alpha & = & L \delta_\alpha m_\beta n^\beta, \label{sf9}\\
x_\alpha & = & - L \epsilon_\alpha m_\beta n^\beta.\label{sf10}
\eeqs
The F-terms setting $\epsilon_\alpha$ and $\delta_\alpha$ to zero
are clearly the ones violating the condition $\det \mesm =\Lambda^8$.
The constraint $m_\alpha n^\alpha \epsilon_\beta \delta^\beta =-\Lambda^8$
will thus send $m_\alpha n^\alpha$ to infinity. Actually we will see
that this scaling to infinity is subdominant with respect to the
one of $m_\alpha$ and $n_\alpha$. For concreteness, let us take
$\epsilon_1=\delta_2=0$ and $\epsilon_2 = \delta_1=\epsilon$.
As an immediate consequence, $x_1= w_2=0$. Moreover, 
\beq 
m_\alpha n^\alpha = {\Lambda^8 \over \epsilon^2}, 
\eeq
so that
\beq
y_3 u_\alpha = L n_\alpha \epsilon^2, \quad
v_\alpha = L m_\alpha \epsilon^2, \quad
w_1 = L {\Lambda^8 \over \epsilon}, \quad
x_2 = -L {\Lambda^8 \over \epsilon}.
\eeq
Comparing with other F-terms, we see that $y_1 \propto y_3$ and
$v_\alpha \propto m_\alpha$. Analyzing the scaling of the basic
loops (\ref{basicloops}), we see that it is consistent
to take the same scaling for $y, m, v$ and $u$.\footnote{Actually, 
the D-flatness conditions will be satisfied
only if we take the elementary fields to scale as above.} 
In this way, 
all loops will scale in the same way and modding out by the
(possibly infinite) common factor we would obtain finite equations.
We thus take
\beq
y_1=y_3=y, \qquad v_\alpha= m_\alpha.
\eeq
We see that this implies
\beq
L={1\over \epsilon^2}, \qquad w_1 = {\Lambda^8 \over \epsilon^3}, \qquad
x_2 = - {\Lambda^8 \over \epsilon^3}.
\eeq
The Lagrange multiplier $L$ goes to infinity, meaning that the mesonic
branch analyzed here is effectively very far from the baryonic branch
described previously. Note that the scaling to infinity behavior
of $w_1$ and $x_2$ is related to the (additional) runaway behavior
of the left-over triangle quiver at any point of the mesonic branch.
Indeed, in the present case, the $SU(3M)$ node at low energies is embedded also
in the $SU(4M)$ of the original quiver, and thus the latter's
dynamical scale is also responsible for this ``secondary'' runaway
behavior. 

The most obvious way to satisfy the F-terms $m_\alpha u^\alpha=0$
would be to take $m_\alpha = u_\alpha$. But that would contradict
the constraint. Hence, we must add a subdominant piece as 
$m_\alpha = u_\alpha+m'_\alpha$, so that 
\beq
m'_\alpha u^\alpha \rightarrow 0 \qquad \mbox{but} \qquad 
y m'_\alpha u^\alpha = {\Lambda^8 \over \epsilon^2}.
\eeq
At this stage, there is some arbitrariness in the way we choose the
scaling to zero of $m'_\alpha u^\alpha$. 

For definiteness, we choose all the non vanishing F-terms to scale
in the same way. Hence we take
\beq
m'_\alpha u^\alpha = \mathcal{O}(\epsilon).
\eeq
This implies the following scaling for $y$
\beq
y = \mathcal{O}(\epsilon^{-3}).
\eeq
As stated previously, we also take $u_\alpha$ to scale in the same way,
$u_\alpha = \mathcal{O}(\epsilon^{-3})$. 
As a consequence, $m'_\alpha= \mathcal{O}(\epsilon^{4})$. 
We can see that all the F-terms,
and the constraint, are satisfied as $\epsilon \rightarrow 0$.\footnote{ 
We are making the reasonable assumption that for large VEVs, 
the K\"ahler potential is close to being the classical canonical one.
Hence also the vacuum energy goes to zero.}
All
the loops have a dominant piece which scales as $\mathcal{O}(\epsilon^{-9})$.
They can actually all be expressed in terms of 3 variables, $y, u_1$ and $u_2$,
so that
\beq
a_\alpha  =y^2 u_\alpha, \quad
b_{\alpha\beta} = y u_\alpha u_\beta, \qquad
c_{\alpha\beta\gamma} = u_\alpha u_\beta u_\gamma.
\eeq
This just reproduces the fact that, away
from the singularity, the space probed by the D3-branes is locally 
$\mathbb{C}^3$ and thus the 20 equations defining the complex cone
can be solved in terms of three complex variables.
Note that we alternatively call $b_{11}\equiv b_1, b_{12}\equiv b_2,
b_{22}\equiv b_3$ and similarly for the $c$s.

\subsection{Recovering the first order complex deformation}
Let us see in more detail how the subdominant piece in
$m_\alpha$ will come into the game, as an ambiguity in defining
the variables with mixed $SU(2)$ indices, i.e. $b_{12}$, $c_{112}$
and $c_{122}$. For instance we can define
\beq
\eta \equiv b_{21}-b_{12}= Y_3 U_{L}^\alpha \mathcal{M}_\alpha
= y u^\alpha {m'}_\alpha = \mathcal{O}(\epsilon^{-2}).
\eeq
The ambiguity increases as $\epsilon \rightarrow 0$, but is vastly
subdominant with respect to the leading behavior of $b$.
Hence at infinity one finds back the algebraic description 
(\ref{algdescrdP1}). This shows that, indeed, one can have a supersymmetric 
configuration on the mesonic branch corresponding to a D3-brane at infinity.

There is a first order complex deformation of the first del Pezzo cone, 
which was given by Altmann \cite{Altmann1} (see also \cite{bhop}) :
\beq
\begin{array}{llll}
a_1 (b_2-3\sigma) = a_2 b_1 &
a_1 b_3 = a_2 b_2 &
b_2 (b_2-3\sigma) = b_1 b_3 &
b_2(b_2-2\sigma) = a_1 c_3 \\
b_2(b_2-4\sigma) = a_2 c_2 &
b_1^2 = a_1 c_1 &
b_3^2 = a_2 c_4 &
a_1 c_2 = b_1 (b_2-\sigma) \\
a_1 c_4 = b_2 b_3 &
a_2 c_1 = b_1 (b_2-3\sigma) &
a_2 c_3 = (b_2-2\sigma) b_3 &
b_1 c_2 = (b_2-\sigma) c_1 \\
b_1 c_3 = (b_2-\sigma) c_2 &
b_1 c_4 = (b_2-\sigma) c_3 &
(b_2-2\sigma) c_3 = b_3 c_2 &
(b_2-2\sigma) c_4 = b_3 c_3 \\
(b_2-2\sigma) c_2 = b_3 c_1 &
c_1 c_4 = c_2 c_3 &
c_2^2 = c_1 c_3 &
c_3^2 = c_2 c_4 
\end{array} \label{defalt}
\eeq
It is natural to ask whether there is a relation between our 
ambiguity parameter $\eta$ and this deformation parameter $\sigma$, 
which we recall has to satisfy $\sigma^2=0$ for consistency.  
We note here that
Altmann's deformation only affects the relations where $b_2 \equiv b_{12}$
appears. In our case, also $c_2$ and $c_3$ would likely be affected.

In order to take into account the ambiguity, we give a more general
definition of the loop variables, keeping the distinction between
$u_\alpha$ and $m_\alpha$. It reads as follows
\begin{eqnarray}\label{abcnew}
a_1 = y^2 u_1,  \quad & b_1 = y u_1 m_1,\quad & c_1=u_1 m_1^2, \nonumber \\
a_2 = y^2 u_2, \quad & b_2 = y u_1 m_2,\quad & c_2=u_1 m_1 m_2,\nonumber \\
& b_3 = y u_2 m_2,\quad & c_3= u_1 m_2^2, \nonumber \\
& & c_4= u_2 m_2^2. 
\end{eqnarray}
The ambiguity will arise when terms like $b_{21}$ appear. It is taken care
of by defining
\beq
u_2 m_1= u_1 m_2 + \eta'.
\eeq
We can now use the above definitions to write how the relations 
(\ref{algdescrdP1}) are deformed
\beq \label{defint}
\begin{array}{cccc}
a_1 (b_2+\eta) = a_2 b_1 &
a_1 b_3 = a_2 b_2 &
b_2 (b_2+\eta) = b_1 b_3 &
b_2^2 = a_1 c_3 \\
b_2 (b_2+\eta) = a_2 c_2 &
b_1^2 = a_1 c_1 &
b_3^2 = a_2 c_4 &
a_1 c_2 = b_1 b_2 \\
a_1 c_4 = b_2 b_3 &
a_2 c_1 = b_1 (b_2+\eta) &
a_2 c_3 = b_2 b_3 &
b_1 c_2 = b_2 c_1 \\
b_1 c_3 = b_2 c_2 &
b_1 c_4 = (b_2+\eta) c_3 &
(b_2+\eta) c_3 = b_3 c_2 &
b_2 c_4 = b_3 c_3 \\
(b_2+\eta) c_2 = b_3 c_1 &
c_1 c_4 = c_2 (c_3 +\tilde \eta) &
c_2^2 = c_1 c_3 &
c_3 (c_3 +\tilde \eta) = c_2 c_4 
\end{array}
\eeq
where we have defined $\tilde \eta = m_2 \eta'$. As we had anticipated, 
some relations involving only $c$s are also deformed, in contradistinction
with (\ref{defalt}). However, it is possible to shift the $c_2$ and $c_3$
variables in such a way that the last three relations above are
not deformed. This is realized by
\beq
c_2 = c_2' -{1\over 3} m_1 \eta' , \qquad c_3 = c_3' - {2\over 3}
m_2 \eta'.
\eeq
As with Altmann's deformation, we are here only considering the
first order deformations, that is we formally impose ${\eta'}^2=0$.

Using now the shifted variables above, we can rewrite all the relations
(\ref{defint}). For instance, take the upper right one, $b_2^2= a_1 c_3$.
In terms of the shifted variables it reads
\beq
b_2 (b_2 + {2\over 3} \eta) = a_1 c_3',
\eeq
so that after identifying 
\beq
\eta \equiv -3\sigma,
\eeq
we recover exactly the right deformed equation as in (\ref{defalt}).
Performing the same shifts in the other relations we recover exactly, 
including all numerical factors, the deformations found by Altmann.

We thus see that regular D3-branes probing the geometry not only
know about the singular cone, but also about its first order
complex deformation. It is because the deformation is only supersymmetric at
first order that the branes are pushed to infinity on the mesonic branch.

We note here that a supergravity approach to deforming the cone over $dP_1$
in the gauge/gravity context has appeared in \cite{blmp}. It is not
immediately clear whether the first order deformation discussed there
exactly maps to Altmann's, described by $\sigma$ above.
It would be very interesting to understand how the deformation
of \cite{blmp} translates into the equations defining the CY cone.

\subsection{Runaway in the $N\neq M$ cases}
Having understood in detail the previous case, we can work out in all
generality the case $N\neq M$ along very similar lines.
As in the conifold case treated in Appendix A, 
we assume that the effect of the quantum dynamics
is to produce an effective ADS-like term in the superpotential. This is 
really the ADS superpotential generated by instanton or gaugino condensation
effects when $N<M$. For $N>M$, the term can be thought of as the result
of integrating out the magnetic quarks when the mesons have (large) VEVs.
In any event, the form of the potential is completely fixed, up to a 
numerical factor, by the symmetries of the problem. Hence, we write
\beq
W= h\mathrm{Tr}(\epsilon^{\alpha\beta} Y_3 U_{L\alpha}\mathcal{M}_{\beta}  
- \epsilon^{\alpha\beta} V_{\alpha}\mathcal{N}_{\beta} + 
\epsilon^{\alpha\beta} V_{\alpha} U_{L\beta}Y_1) +(M-N)\left( 
{\Lambda_2^{N+7M}\over \det \mesm} \right)^{1\over M-N}.
\eeq
The F-term equations derived from the superpotential above will be 
much similar as before. The eqs. (\ref{ef5})--(\ref{ef7}) remain unchanged, 
while the eqs. (\ref{ef3})--(\ref{ef4}) become
\beqs
\epsilon^{\alpha\beta}  Y_3 U_{L\alpha}  & = & 
\left( {\Lambda_2^{N+7M}\over \det \mesm} \right)^{1\over M-N}
{1\over \det \mesm}
{\partial \det{\mesm} \over \partial \mathcal{M}_{\beta}}, \label{ef3new}\\
\epsilon^{\alpha\beta} V_{\alpha} & = & -
\left( {\Lambda_2^{N+7M}\over \det \mesm} \right)^{1\over M-N}
{1\over \det \mesm}
{\partial \det{\mesm} \over \partial \mathcal{N}_{\beta}}. \label{ef4new}
\eeqs
Of course, the F-terms involving the baryons are no longer present. 
In the following it will be convenient to introduce the shorthand
\beq
\mathcal{L} \equiv 
\left( {\Lambda_2^{N+7M}\over \det \mesm} \right)^{1\over M-N}.
\eeq
We can now attempt to solve the F-terms using an ansatz exactly similar
to (\ref{ans1})--(\ref{ans3}), except that now $m_\alpha, n_\alpha,
y_3, v_\alpha, y_1$ and $u_\alpha$ are $N\times N$ diagonal matrices,
while $\epsilon_\alpha, \delta_\alpha, w_\alpha$ and $x_\alpha$ are
$M\times M$ diagonal matrices. 

We can further simplify the problem by taking all the matrices to be 
proportional to the identity. Of course, as far as the $N\times N$ matrices
are concerned, we really want ultimately all the eigenvalues to be distinct, 
but the scalings discussed below will not change.

Thus, introducing the ansatz above in the F-term equations, we will obtain
simplified equations which consist of (\ref{sf1})--(\ref{sf6})
together with (up to an $N, M$-dependent sign)
\beqs
y_3 u_\alpha &= & \mathcal{L} {1\over m_\beta n^\beta} n_\alpha, 
\label{sf7new}\\
v_\alpha &= & \mathcal{L} {1\over m_\beta n^\beta} m_\alpha, \label{sf8new}\\
w_\alpha & = & -\mathcal{L} {1\over \epsilon_\beta \delta^\beta}\delta_\alpha, 
\label{sf9new}\\
x_\alpha & = & \mathcal{L} {1\over \epsilon_\beta \delta^\beta}
\epsilon_\alpha.\label{sf10new}
\eeqs
We then again take $\epsilon_1=\delta_2=0$ and 
$\epsilon_2 = \delta_1=\epsilon$. This implies $x_1= w_2=0$.
We further simplify and solve more F-terms by taking 
$y_1=y_3=y$, $v_\alpha=m_\alpha$ and $n_\alpha= yu_\alpha$.
Then the F-terms (\ref{sf7new})--(\ref{sf8new}) are solved by
\beq
\mathcal{L} = y m_\alpha u^\alpha. \label{scalel}
\eeq
We eventually recover as before that all the F-terms are satisfied
if we take $\epsilon \rightarrow 0 $ together with 
\beq
m_\alpha u^\alpha = \mathcal{O}(\epsilon).
\eeq
This again implies that there is a subleading component in $m_\alpha$,
and we made the (arbitrary) choice of scaling all the non-vanishing
F-terms to zero in the same way.

The scaling (to infinity) of $y$ is determined in the following way. 
Eq. (\ref{scalel}) becomes now $\mathcal{L} = y\mathcal{O}(\epsilon)$.
However $\mathcal{L}$ is an expression involving $y$ and $\epsilon$.
Indeed, up to a sign 
\beq
\det \mesm = y^N (m_\alpha u^\alpha)^N \epsilon^{2M} \sim
y^N \epsilon^{N+2M}.
\eeq
It is then easy to see that $\mathcal{L} \sim y\epsilon$
implies
\beq
y=\mathcal{O}(\epsilon^{-3}).
\eeq
We can then also take $u_\alpha$ to scale in the same way. 
Thus, all the scalings
are exactly the same as in the previous simple case of $N=M=1$, and we
are led to the same conclusions regarding the asymptotic behavior
of the loop variables and their ambiguities.

In particular, the ambiguity parameter $\eta$ which is eventually
equated to the first order deformation parameter, is directly
proportional to $\mathcal{L}$, which in turn is proportional to the
gaugino condensate $S$ for the second node (strictly speaking, in the
case $N<M$).

Note that with the scalings above, $\mathcal{L} = \mathcal{O}(\epsilon^{-2})$.
Thus it scales like the Lagrange multiplier in the previous case, and actually
one can show that it is indeed formally replaced by the Lagrange
multiplier when $N=M$ in all generality.
This scaling also implies that the gaugino condensate grows unboundedly along
the runaway direction.

As a last curiosity, we can compute how the determinant scales
\beq
\det \mesm \sim \epsilon^{2(M-N)}.
\eeq
Quite intuitively, the determinant goes to zero when it has a predominance
of zero eigenvalues (in the $N<M$ case) while it goes to infinity
when there are more eigenvalues scaling to infinity (when $N>M$).

The global picture after this analysis is the following.
If we start with a number of regular branes $N$ much larger than the
number of fractional ones $M$, we see that we always have the option 
of trying to explore the mesonic moduli space, which is represented
by the regular branes wandering around the geometry. However, 
this mesonic moduli space, for any value of $N$, is actually lifted
because of the presence of the fractional branes, and the regular
branes are pushed to infinity (as anticipated in \cite{fhsu} for the case of
one probe regular D3-brane), where they explore a geometry very close
to the singular complex cone over $dP_1$. This runaway behavior seems
to have the same strength irrespective of the relative numbers
$N$ and $M$. However we recall that there was some freedom to choose
the scaling of the variables, so that this dynamical issue is not settled
at this level of the analysis.

At any given $N$, one has however also the option of exploring the
``baryonic'' branch of the moduli space, which for $N>M$ amounts
to performing a Seiberg duality on the second node. This restitutes
the same quiver but with ranks shifted according to $N \rightarrow N-M$.
Then at any further step one finds again the alternative between
going on a runaway mesonic branch or performing a further step.
At the last step, we either end up with a runaway mesonic branch
at $N<M$, or if $N=M$, we have a last option of going toward a
baryonic branch, which however is itself runaway (albeit
differently) as was already known. Note that in this last case, 
we do not have a pictorial way of representing the runaway as
some branes being pushed to infinity. It would be nice to understand
this better.

\section{Discussion}
It was suggested in \cite{bhop} that there was a one to one correspondence
between CY singularities with obstructed deformations and quiver gauge
theories with runaway supersymmetry breaking in the deep IR. In the present
paper we have shown that, in the example of the $dP_1$ geometry,
the relation is even more precise. At higher steps of the cascade,
there is also a runaway behavior along the mesonic branches, which reproduce
exactly the first order deformation of the geometry.
Given the genericity of runaway behavior in quiver gauge theories
(see e.g.~\cite{Brini,Butti}), 
we expect that obstructed deformations can be reproduced similarly in generic
toric singularities, even when non-obstructed deformations are possible.
Indeed, in Appendix B we show that this is true for
the $dP_2$ singularity.

It is nice to see that the correspondence between quiver gauge theories
and D-branes at singularities remains valid 
beyond issues pertaining to supersymmetric vacua. This was also argued to hold
for theories displaying metastable vacua (see \cite{abfk} for examples where
metastability can be argued for on both sides of the correspondence). 
The situation discussed here is
qualitatively different and can thus be considered as further evidence.

On the mesonic branches, the runaway is naturally interpreted as
D3-branes being pushed to infinity.
Unfortunately we do not have as nice an interpretation of the
runaway along the baryonic branch. 
As argued in \cite{fhsu}, it could be related
to the blowing up of a closed string modulus, namely a dynamical
FI term. This blown-up background must somehow have an unbalanced
D3-charge/tension ratio, so that the additional regular D3 branes
feel a repulsive force in its presence.
Based on the findings presented here, it is tempting to speculate that
if a supergravity dual of the baryonic branch runaway exists, a crucial
role in its construction should be played by a (non-supersymmetric) completion
of the first order deformation of the $dP_1$ cone. Moreover, there should
be a signal of a diverging gluino condensate. Presumably, a singularity
is impossible to avoid, at least in a static solution.

Having many runaway directions might eventually be interesting 
in cosmology, which is the only framework to make sense of such
theories with no vacuum. In particular, there can be different
regions of the universe where the runaway is taking place along
a different direction. There would then be domain walls between
those regions (possibly bubble walls if the runaway is faster
in some specific direction). Those will be NS5-branes wrapped
on the topological $S^3$ of the base. This can be seen using the
same arguments as in \cite{kpv,ekk} and noticing that the domain walls
would interpolate between regions with a different number of D3-branes.
Note that for the domain wall tension to be non vanishing, 
the 3-cycle wrapped by the NS5-branes must be of finite size.
This is non trivial in the absence of a consistent deformation
(i.e.~a blown up 3-cycle). We are left to suppose that
the 3-form flux sourced by the fractional branes somehow
prevents the collapse of the NS5-brane worldvolume, possibly
due to dynamics which is necessarily time-dependent.

\subsection*{Acknowledgements}

We would like to thank F.~Bigazzi for collaboration at the
early stages of this work, and M.~Bertolini, B.~Craps, S.~Cremonesi,
J.~Evslin, C.~Krishnan,
S.~Kuperstein and S.~Pinansky for interesting discussions and correspondence.
This work is partially supported by the European Commission FP6
Programme MRTN-CT-2004-005104, in which the authors are associated to
V.U. Brussel, by IISN - Belgium 
(convention 4.4505.86) and by the 
``Interuniversity Attraction Poles Programme --Belgian Science Policy''. 
R.A. is a Research Associate of the Fonds National de la Recherche
Scientifique (Belgium). C.C. is a Boursier FRIA-FNRS.

\appendix

\section{Classical and quantum moduli spaces of the conifold theory}
We consider here the warm-up example of the quiver gauge theory
resulting from $N$ D3-branes at a conifold singularity, with 
the addition of $M$ fractional branes. The gauge groups are $SU(N)\times
SU(N+M)$, and there are two pairs of bifundamentals $A^i_{\alpha a}$
and $B^a_{\alpha i}$ where $\alpha=1,2$ and $i$ and $a$ are indices in the
first and second gauge group respectively. We aim here at reproducing
in a compact way the results of \cite{dks}.

\subsection{Classical analysis}
The classical tree level superpotential is 
\beq
W= h A^i_{\alpha a} B^a_{\beta j} A^j_{\gamma b} B^b_{\delta i}
\epsilon^{\alpha\gamma} \epsilon^{\beta\delta}.
\eeq
From it we derive the F-terms
\beq
A^i_{\alpha a} B^a_{\beta j} A^j_{\gamma b} \epsilon^{\alpha\gamma}=0, \qquad
B^a_{\beta j} A^j_{\gamma b} B^b_{\delta i}\epsilon^{\beta\delta}=0.
\eeq
The above F-terms can be written in a more interesting way if contracted
so as to form gauge invariants of the first or the second gauge group.
We call $M_{\alpha\beta j}^i= A^i_{\alpha a}B^a_{\beta j}$ and 
$\tilde M_{\alpha\beta b}^a=B^a_{\alpha i}A^i_{\beta b}$, and we obtain,
in matrix notation
\beq
M_{\alpha\beta} M_{\gamma\delta}\epsilon^{\alpha\gamma}=0=
M_{\alpha\beta} M_{\gamma\delta}\epsilon^{\beta\delta}, \qquad
\tilde M_{\alpha\beta} \tilde M_{\gamma\delta}\epsilon^{\alpha\gamma}=0=
\tilde M_{\alpha\beta} \tilde M_{\gamma\delta}\epsilon^{\beta\delta}.
\eeq
The above equations read, component by component
\beqs
M_{11}M_{21}=M_{21}M_{11},& \qquad& M_{11}M_{22}=M_{21}M_{12}, \nonumber \\
M_{12}M_{21}=M_{22}M_{11}, & \qquad&  M_{12}M_{22}=M_{22}M_{12},\nonumber \\
M_{11}M_{12}=M_{12}M_{11},& \qquad& M_{11}M_{22}=M_{12}M_{21}, \nonumber\\
M_{21}M_{12}=M_{22}M_{11}, & \qquad&  M_{21}M_{22}=M_{22}M_{21},
\eeqs
or, in a more compact way
\beq
[M_{\alpha\beta},M_{\gamma\delta}]=0, \qquad M_{11}M_{22}-M_{12}M_{21}=0.
\eeq
The same holds for the matrices $\tilde M_{\alpha\beta}$.
As a consequence, using gauge transformations of $SU(N)$ and $SU(N+M)$
respectively, one can diagonalize both sets of 4 commuting matrices
$M_{\alpha\beta}$ and $\tilde M_{\alpha\beta}$. Note that the latter matrices
are not of maximal rank $N+M$, but rather only of rank $N$. Hence, they
will have generically $M$ vanishing eigenvalues.

Eigenvalue by eigenvalue, we have that
\beq
 m^{(i)}_{11} m^{(i)}_{22} - m^{(i)}_{12} m^{(i)}_{21}=0.
\label{conif}
\eeq
These are $N$ copies of the equation defining the conifold singularity,
$xy=uv$. If we define 
\beq
{\mathcal{M}} \equiv \left( \begin{array}{cc} M_{11} & M_{12} \\
M_{21} & M_{22} \end{array}\right),
\eeq
we immediately see that 
\beq
\det {\mathcal{M}} = 0.
\eeq
As for $\tilde M_{\alpha\beta}$, we can take the first $N$ entries of, say,
$\tilde M_{11}$ to be non vanishing. Then the relations similar
to (\ref{conif}) are most generically satisfied by non vanishing eignevalues
when also the other 3 matrices have non vanishing first $N$ entries.

At this stage, let us go back to the F-term conditions written in terms of
the elementary fields. For instance, we have the following expression
(even before imposing the F-terms)
\beq
M_{11} A_1 = A_1 B_1 A_1 = A_1 \tilde M_{11}.
\eeq
For a generic matrix $A_1$ and $M_{11}$, $\tilde M_{11}$ as above,
we have
\beq
m^{(i)}_{11} A^i_{1a}= A^i_{1a} \tilde m^{(a)}_{11}.
\eeq
As we need some components of $A_1$ to be non zero (since after all
$M_{11}$ and $\tilde M_{11}$ are built from it), we see that
we must have $m^{(i)}_{11}=\tilde m^{(i)}_{11}$ and
$A^i_{1a}=0 $ for $i\neq a$.

Now, using the F-terms we also obtain that
\beq
M_{12}A_1=A_1 \tilde M_{21}, \quad
M_{21}A_1= A_1 \tilde M_{12}, \quad
M_{22}A_1= A_1 \tilde M_{22},
\eeq
so that 
\beq
m^{(i)}_{12}=\tilde m^{(i)}_{21}, \quad
m^{(i)}_{21}=\tilde m^{(i)}_{12}, \quad
m^{(i)}_{22}=\tilde m^{(i)}_{22}.
\eeq
Finally, using relations based on $A_2B_2A_2$, $B_1A_1B_1$ and $B_2A_2B_2$
we obtain that all elementary fields $A_\alpha$, $B_\alpha$
can be taken to be diagonal in their upper/left $N\times N$ piece.

Note that we did not use until now information coming from requiring
D-flatness. The only constraint left is
\beq
| a^{(i)}_1|^2+| a^{(i)}_2|^2 = | b^{(i)}_1|^2+| b^{(i)}_2|^2.
\eeq

\subsection{Quantum analysis}
We now want to take into account quantum corrections to the above story.
We do this by considering that the node with largest rank $SU(N+M)$
goes to strong coupling first. Then, its dynamics should be effectively
described by gauge invariants, which in this case are the $M_{\alpha\beta}$
matrices with indices in the first gauge group, which will be considered
as classical in these considerations.

The quantum corrections in a region of the moduli space where the 
mesons $M_{\alpha\beta}$ have large enough VEVs (i.e. the so-called
mesonic branch) are captured by adding an ADS-like superpotential.
It can be seen to arise in the Seiberg dual picture 
from integrating out the dual magnetic quarks
which are massive because of the mesonic VEVs. We thus write
\beq
W_\mathrm{eff} = h M_{\alpha\beta} M_{\gamma\delta} 
\epsilon^{\alpha\gamma} \epsilon^{\beta\delta} 
- (N-M) \left( {\Lambda^{N+3M} \over \det \mathcal{M}}\right)^{1\over M-N},
\eeq
where $\Lambda$ is the dynamical scale of the strongly coupled node.
Note that for $N<M$, this is really an ADS superpotential. For $N>M$,
the determinant has actually a positive power. The case $N=M$ is analyzed
below in more detail.

Extremizing with respect to $M_{\alpha\beta}$, we obtain
\beq
h \left( \begin{array}{cc} M_{22} & -M_{12} \\
-M_{21} & M_{11} \end{array}\right) = 
\left( {\Lambda^{N+3M} \over \det \mathcal{M}}\right)^{1\over M-N}
\mathcal{M}^{-1}. \label{feff}
\eeq
Multiplying by $\mathcal{M}$ these equations from the right and from the
left, we obtain matrix equations which imply
\beq
[M_{\alpha\beta},M_{\gamma\delta}]=0, \qquad M_{11}M_{22}-M_{12}M_{21}=
{1\over h}\left( {\Lambda^{N+3M} \over \det \mathcal{M}}\right)^{1\over M-N}.
\eeq
As before, the matrices $M_{\alpha\beta}$ can all be simultaneously
diagonalized, and their eigenvalues must satisfy
\beq
m^{(i)}_{11} m^{(i)}_{22} - m^{(i)}_{12} m^{(i)}_{21}={1\over h}\left(
{\Lambda^{N+3M} \over \prod_j (m^{(j)}_{11} m^{(j)}_{22} 
- m^{(j)}_{12} m^{(j)}_{21})}
\right)^{1\over M-N}. \label{defcon}
\eeq
Taking the product of all the $N$ such equations, we eventually obtain
\beq
\det \mathcal{M} = \prod_i (m^{(i)}_{11} m^{(i)}_{22} - 
m^{(i)}_{12} m^{(i)}_{21})= \left( h^{N-M}\Lambda^{N+3M}\right)^{N\over M}.
\eeq
Reinserting in (\ref{defcon}), we obtain
\beq
m^{(i)}_{11} m^{(i)}_{22} - m^{(i)}_{12} m^{(i)}_{21}=
\left( h^{N-M}\Lambda^{N+3M}\right)^{1\over M}=\Lambda^4 
(h\Lambda)^{N-M\over M}.
\eeq
We thus see that we have $N$ copies of the deformed conifold, defined
by $xy-uv=\epsilon$. Note that the deformation parameter is parametrically
smaller as $N$ is increased, since $h$ can be taken to be of the order
of the inverse string scale.\footnote{The string scale is effectively
warped down from its true value in the deep UV. This can be seen as
an effect of the cascading RG flow.}

Thus we see that when fractional branes are present, the moduli space probed
by regular branes is smoothened to the
deformed conifold because of quantum effects.

When there are no fractional branes, $M=0$, the equations (\ref{defcon})
can be satisfied only if $\det \mathcal{M}=0$, which implies eventually
(\ref{conif}), i.e. the moduli space remains the classical, singular
conifold. 

Note that there is a subtle point in this specific case. The F-terms
(\ref{feff}) would seem
to imply that the mesons actually have to vanish. This is clearly a too
strong constraint. Hence, requiring F-flatness 
of the effective superpotential in this case seems
to be misleading. Possibly, this is due to the strictly conformal nature
of the quiver gauge theory, which prevents us to consider one
node as strongly coupled and the other as classical.

\subsection{When baryonic branches are present}
We are left to analyze the case $N=M$, which we take to be a case study
of the case $N=kM$ where baryonic operators are allowed.
At the classical level, we can write two more gauge invariants
of the second node, which turn out to be gauge invariant also with respect
to the first one. Indeed, node two has $N_f=N_c$ and we can write
\beqs
\mathcal{B} &= &\epsilon_{i_1\dots i_{2M}}\epsilon^{a_1\dots a_M}
\epsilon^{b_1\dots b_M} A^{i_1}_{1 a_1}\dots A^{i_M}_{1 a_M}
A^{i_{M+1}}_{2 b_1}\dots A^{i_{2M}}_{2 b_M}, \nonumber \\
\tilde \mathcal{B} &= &\epsilon^{i_1\dots i_{2M}}\epsilon_{a_1\dots a_M}
\epsilon_{b_1\dots b_M} B^{a_1}_{1 i_1}\dots B^{a_M}_{1 i_M}
B^{b_1}_{2 i_{M+1}}\dots B^{b_M}_{2 i_{2M}}.
\eeqs
Still at the classical level, we see that we can form new gauge invariants
from the F-terms such as
\beq
M_{\alpha\beta} \mathcal{B}  = 0 , \qquad M_{\alpha\beta}
\tilde \mathcal{B}  = 0.
\eeq
It implies that we can turn on either the baryonic VEVs or the mesonic
ones, but not both at the same time.
Moreover, the classical constraint $\det \mathcal{M}=  \mathcal{B}
\tilde \mathcal{B}$ implies that $\det \mathcal{M}=0$ on the mesonic branch
(this was already derived above) and that $\mathcal{B}\tilde \mathcal{B}=0$
on the baryonic branch, which is thus separated in two components.

The mesonic branch is derived exactly as before, so that the complete
moduli space in this case is the sum of the symmetric product of $M$ copies
of the conifold (parametrized by $M_{\alpha\beta}$) and two complex lines
(parametrized by $\mathcal{B}$ and $\tilde \mathcal{B}$). 
All three components of the moduli
space meet at the origin of each branch.

At the quantum level, the effective strongly coupled 
dynamics of the second node induces a deformation of its classical moduli
space. Such a deformation is encoded in the following effective
superpotential which includes a Lagrange multiplier $L$
\beq
W_\mathrm{eff} = h M_{\alpha\beta} M_{\gamma\delta} 
\epsilon^{\alpha\gamma} \epsilon^{\beta\delta} + L(\det\mathcal{M}
- \mathcal{B}\tilde \mathcal{B}- \Lambda^{4M}).
\eeq
The F-terms are the following
\beq
h \left( \begin{array}{cc} M_{22} & -M_{12} \\
-M_{21} & M_{11} \end{array}\right) = L (\det\mathcal{M})\mathcal{M}^{-1},
\eeq
\beq
L \mathcal{B}=0=L \tilde \mathcal{B},
\eeq
together with the constraint
\beq
\det\mathcal{M}- \mathcal{B}\tilde \mathcal{B}=\Lambda^{4M}.
\eeq
It is clear that we have a baryonic branch where the $\mathcal{B},
\tilde \mathcal{B}\neq 0$. This implies that $L=0$ and in turn
$M_{\alpha\beta}=0$. The two classical baryonic branches have merged into
one $\mathcal{B}\tilde \mathcal{B}=-\Lambda^{4M}$.

If we want the mesons to be non vanishing, we need to have $L\neq 0$,
which forces the baryons to vanish. Then we automatically get 
$\det\mathcal{M} =\Lambda^{4M}$, the $M_{\alpha\beta}$ commute
and eigenvalue by eigenvalue we have
\beq
m^{(i)}_{11} m^{(i)}_{22} - m^{(i)}_{12} m^{(i)}_{21}=\Lambda^4, 
\eeq
which also sets $L=h \Lambda^{4-4M}$.

Thus we see that at the quantum level, we still have two components,
one being the one complex dimensional baryonic branch and the other
being the symmetric product of $M$ copies of the deformed conifold.
This time the two branches are both completely smooth\footnote{Except
of course for singularities due to the symmetric product orbifold action.} 
and do not touch. 

The full moduli space of the theory for a given $N$, can be derived
component by component in the way described here, reproducing the results
of \cite{dks}. The first component
is described by $N$ copies of the deformed conifold, corresponding
to the mesonic branch. However if the mesons are not given VEVs, one
can Seiberg dualize the strongly coupled node and reach a theory
where effectively $N$ is replaced by $N-M$. At every step in this cascade
of dualities there is a component of the moduli space which will be
described by $N-kM$ copies of the deformed conifold. If $N$ is a multiple
of $M$, we end up with a smooth baryonic branch, while if it is not
the smallest component of the moduli space will still be a mesonic branch
corresponding to $N-k_\mathrm{max}M<M$ D3 branes on the deformed conifold.

\section{Quantum corrections to the $dP_2$ moduli space}

\subsection{The  space of complex deformations for $dP_2$}
The complex cone over $dP_2$ admits two different kinds of fractional branes, according to the classification of \cite{fhsu}. One is a deformation brane, corresponding to a complex structure deformation of the cone. The corresponding gauge theory was studied in detail in \cite{Pinansky}, where it was shown that the deformed chiral algebra encodes precisely the complex deformation computed according to Altmann's rules \cite{Altmann1}. The second fractional brane allowed by the geometry is a so called supersymmetry breaking (SB) brane, which corresponds in this case to an obstructed complex deformation of the geometry.

This section follows closely the work of \cite{Pinansky}, to which we refer for further details on the application of Altmann's techniques. Using the usual toric geometry techniques, one can describe the $dP_2$ cone as an affine variety in $\mathbb{C}^8$. Let  $(a_1,a_2,b_1,b_2,b_3,c_1,c_2,d)\in \mathbb{C}^8$ be the complex coordinates corresponding to the generators of the dual toric cone $\sigma^{\vee}$.  There are 14 relations amongst these:
\beqs 
b_2^2=b_1b_3,\quad & b_2^2=a_1c_2,\quad & b_2^2=c_1a_2,\quad \nonumber\\
c_1^2=b_1d,\quad & c_2^2=b_3d,& \quad \nonumber\\
 b_1a_2=b_2a_1,\quad & c_1b_2=c_2b_1,\quad & b_2a_2=b_3a_1,\nonumber\quad\\
 c_1b_3=c_2b_2 \quad &  b_1b_2=c_1a_1,\quad & b_2b_3=c_2a_2,\nonumber\quad\\
c_1c_2=b_2d,\quad & c_1b_2=a_1d,\quad & c_2b_2=a_2d \label{reladP2}. 
\eeqs
The Minkowski cone (that is the cone of Minkowski summands of the toric diagram) is given by 
\beqs
(t_1,...,t_5) \quad \mathrm{s.th.} && \quad P_1(t)=t_1-t_2-t_3+t_5 =0, \nonumber\\&& \quad P_2(t)=t_1+t_2-t_4-t_5=0, 
\eeqs
and $t_i\geq 0$. We parametrise it by
\beq
t_1=t, \quad t_2=t-s_1, \quad t_3=t-s_2, \quad t_4=t+s_2, \quad t_5 = t-s_1-s_2 .
\eeq
The space of deformations is given by imposing the further constraints 
$P_{1,2}(t^2)=0$. We thus have $s_1$ and $s_2$ subject to the quadratic constraints
\beq
s_1s_2=0, \qquad s_2^2=0.
\eeq 
Clearly, one solution is $s_2=0$. This corresponds to the deformation brane case studied in \cite{Pinansky}. But we are interested here in the case $s_1=0$, $s_2^2=0$. Then, $s_2$ corresponds to a first order deformation obstructed at second order (corresponding to the SB branes), similarly to the $dP_1$ case.
Running Altmann's algorithm, which replaces the coordinate $b_2$ by the five new variables $t_i$, 
we can show that the 14 relations (\ref{reladP2}) become
\beqs 
t_1^2=b_1b_3,\quad &  t_2 t_4=a_1c_2,\quad &  t_1 t_2 =c_1a_2,\quad \nonumber\\
c_1^2=b_1d,\quad &   c_2^2=b_3d,& \quad \nonumber\\ 
b_1a_2=\frac{t_1^2}{t_4}a_1,\quad & c_1t_1=c_2b_1,\quad & t_4 a_2=b_3a_1,\quad\nonumber\\
 c_1 b_3=c_2t_1 \quad &  b_1\frac{t_2t_4}{t_1}=c_1a_1,\quad & t_2b_3=c_2a_2,\quad \nonumber\\
c_1c_2=t_1 d,\quad & c_1\frac{t_2t_4}{t_1}=a_1d,\quad & c_2t_2=a_2d . 
\eeqs
Restricting to the case $s_1=0$, $s_2^2=0$, we see that $t_1=t_2$, $t_3=t_5$, and (renaming $t=b_2$, $s_2=\sigma$) we eventually find
\beqs 
b_2^2=b_1b_3,\quad &  b_2 (b_2+\sigma)=a_1c_2,\quad &  b_2^2 =c_1a_2,\quad\nonumber \\
c_1^2=b_1d,\quad &   c_2^2=b_3d,& \quad \nonumber\\ 
b_1a_2=(b_2-\sigma)a_1,\quad &  c_1b_2=c_2b_1,\quad & (b_2+\sigma) a_2=b_3a_1,\quad\nonumber\\
 c_1 b_3=c_2b_2 \quad &  b_1(b_2+\sigma)=c_1a_1,\quad & b_2b_3=c_2a_2,\quad\nonumber\\
c_1c_2=b_2 d,\quad & c_1(b_2+\sigma)=a_1d,\quad & c_2b_2=a_2d \label{fstordP2}. 
\eeqs
Notice that indeed, for consistency, it implies that we must have $\sigma^2=0$.

\subsection{Classical moduli space}
\begin{figure}[h]
\begin{center}
\includegraphics[height=5.5cm]{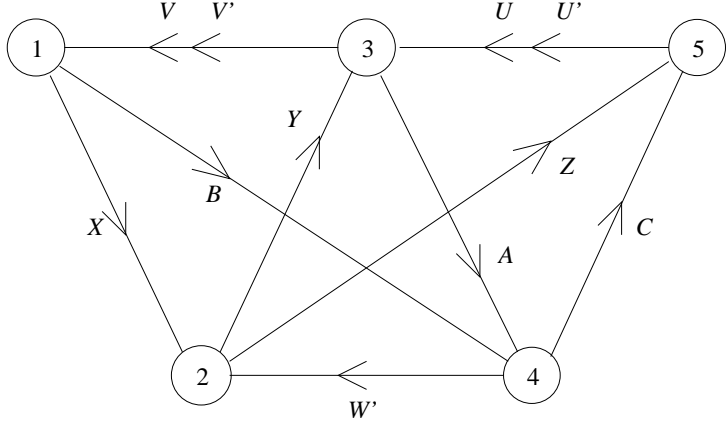}
\end{center} 
\caption{The $dP_2$ quiver.}
\label{fig03}
\end{figure}

When $N$ D3-branes, $M$ SB branes and $P$ deformation branes are present on the $dP_2$ cone, the corresponding gauge theory has an $SU(N+M+P)\times SU(N+2M)\times SU(N+M)\times SU(N)\times SU(N+P)$ gauge group. We study here the  $P=0$, $N=M=1$ case. The field content can be read from the quiver in figure \ref{fig03}.

The tree level superpotential is given by
\beq
W_{tree} = W'YA - XYV' - ACU' + XZU'V + BCUV' -W'ZUVB,
\eeq
where the trace is implied.

Using the F-conditions, there is a minimal set of 8 loops of the quiver that generate all mesonic gauge invariants, the chiral primaries. These are given by
\beqs
a_1 = XYV, \quad & b_1 = BCU'V, &\quad c_1= BCU'V',\nonumber\\
a_2 = XZUV, \quad & b_2 = XZU'V, &\quad c_2= XZU'V',\nonumber\\
 \quad & b_3 = XZUV', &\quad d= BW'ZU'V',\label{loopdP2}
\eeqs
where we chose to base all loops at the first node. Note that there is a grading in term of the number of primed fields. There are 14 relations amongst these 8 fields, defining the complex cone over $dP_2$ as in (\ref{reladP2}).

\subsection{Runaway on the mesonic branch}

We will consider the quantum corrections from the second node to be dominant. It has gauge group $SU(3)$ and $N_f=N_c$. As for $dP_1$, we must consider its mesons as effective fields,
\beqs
M_1= XY, \qquad && M_2= XZ, \\
M_3= W'Y, \qquad && M_4= W'Z. 
\eeqs
Let us also define  the ($3\times 3$) meson matrix,
\beq
{\mesm} \equiv \left( \begin{array}{cc} M_1 & 
M_2 \\
 M_3& M_4  \end{array}\right).
\eeq

The quantum contribution to the superpotential is of course
\beq
W_{qu} = L(\det{\mesm}-\bar \mathcal{B}\bar \mathcal{B} -\Lambda^6_2). 
\eeq
We want to analyse the behavior of the mesonic branch, so we will impose the above constraint as $\det{\mesm} = \Lambda^6_2$. Similarly to the $dP_1$ case, we can take all bifundamental fields to be upper-left diagonal:
\beq \nonumber
M_1 = \left(\begin{array}{cc} m_1 & 0  \\ 0 & \epsilon  \end{array}\right),
\qquad M_2= \left(\begin{array}{c} m_2 \\ 0  \end{array}\right),
\eeq
\beq \nonumber
 M_3= \left(\begin{array}{cc} m_3 & 0  \end{array}\right),
 \qquad M_4= m_4,
\eeq
\beq \nonumber
A =  \left(\begin{array}{c} a \\ 0  \end{array}\right),
\qquad  B =  \left(\begin{array}{c} b \\ 0  \end{array}\right), \qquad C=c,
\eeq
\beq 
U =  \left(\begin{array}{cc} u & 0  \end{array}\right),
\qquad V =  \left(\begin{array}{cc} v & 0  \\0 & 0 \end{array}\right),
\eeq
\beq 
U' =  \left(\begin{array}{cc} u' & 0  \end{array}\right),
\qquad V' =  \left(\begin{array}{cc} v' & 0  \\0 & \tilde{v}' \end{array}\right).
\eeq
The non-upper-left elements $\epsilon$ and $\tilde{v}'$ are of course there
to reconcile the rank condition and the constraint.
 
The constraint is thus
\beq \label{constdP2}
\det{\mesm} = \epsilon (m_1m_4- m_2m_3) = \Lambda^6_2. 
\eeq

With the parametrisation above, the VEVs of the bifundamentals are c-numbers, and the superpotential is given by 
\beqs
W &=& m_3a - m_1v' -\epsilon \tilde{v}' -acu' + m_2u'v + bcuv' - m_4uvb \nonumber \\ &&+ L(\epsilon(m_1m_4- m_2m_3)- \Lambda^6_2).
\eeqs
The F-terms are
\beqs
F_a =m_3 -cu' &=&0, \label{dp2f1}\\
F_b = cuv'-m_4uv &=&0,\\
F_c = -u'a+uv'b&=&0,\\
F_u = v'bc -vbm_4&=&0,\\
F_v = m_2u' -bm_4u&=&0,\\
F_{u'} = -ac+ vm_2&=&0,\\
F_{v'} = -m_1+ bcu&=&0,\\
F_{\tilde{v}'} = -\epsilon &=&0, \label{dp2f8} \\
F_{\epsilon} = -\tilde{v}' + L(m_1m_4- m_2m_3) &=&0 ,\\
F_{m_1} = -v'+ L\epsilon m_4  &=&0,\\
F_{m_2} = u'v- L\epsilon m_3 &=&0, \\
F_{m_3} = a- L\epsilon m_2&=&0,\\
F_{m_1} = -uvb+ L\epsilon m_1 &=&0. 
\eeqs
Clearly, (\ref{dp2f8}) is incompatible with (\ref{constdP2}) unless $(m_1m_4-
m_2m_3) =\frac{\Lambda^6_2}{\epsilon} \rightarrow \infty$. We can solve for
$F_{m_i}=0$ by taking $L=\Lambda^6_2 /\epsilon$ (so that the baryonic branch indeed decouples as $\epsilon$ goes to zero), and 
\beqs
m_1 = uvb, & \qquad & m_2 =a,\nonumber \\
m_3 = u'v & \qquad & m_4 =v'.
\eeqs
Then the other F-terms imply $c=v$. Moreover, (\ref{constdP2}) becomes
\beq
m_1m_4- m_2m_3 = v(ubv'-au') = v F_c = \frac{\Lambda^6_2}{\epsilon}.
\eeq
We can choose $F_c$ to scale to zero as 
\beq\label{Fc0}
F_c = uv'b -u'a = \mathcal{O}(\epsilon),
\eeq
It implies that $v$ must scale as
\beq
v =c = \mathcal{O}(\epsilon^{-2}).
\eeq
One  can then easily check that all F-terms are satisfied, with $F_c=F_v=0$ that must be satisfied in the limit $\epsilon \rightarrow 0$. By taking the simplest solution $v=c$, $u=b$, $a=ub=b^2$, $u'=v'$, one can express (\ref{loopdP2}) in term of $(b,c,u')\in \mathbb{C}^3$ :
\beqs
a_1 = b^2c^2, \quad & b_1 = bc^2u', &\quad c_1= bc(u')^2,\nonumber\\
a_2 = b^3c, \quad & b_2 = b^2cu', &\quad c_2= b^2 (u')^2,\nonumber\\
 \quad & b_3 = b^3u', &\quad d= b (u')^3,\label{c3dP2}
\eeqs
which implies the 14 relations (\ref{reladP2}). Note that all coordinates go to infinity as $\epsilon^{-8}$.

\subsection{Recovering the first order complex deformation}
To conclude, let us show that the gauge theory result reproduces the first order complex deformation (\ref{fstordP2}). The ambiguity coming from solving (\ref{Fc0}) can be accounted for by defining
\beq
v' = u' + \tilde{\eta}.
\eeq
Let us solve for the loops as in (\ref{c3dP2}), but taking this ambiguity into account:
\beqs
a_1 = b^2c^2, \quad & b_1 = bc^2u', &\quad c_1= bc u'v',\nonumber\\
a_2 = b^3c, \quad & b_2 = b^2cu', &\quad c_2= b^2 u'v',\nonumber\\
 \quad & b_3 = b^3 v', &\quad d= b u'(v')^2\label{c3dP2bis}.
\eeqs
Now it is an easy matter to construct the deformed relations amongst the variables of (\ref{c3dP2bis}). We find
\beqs 
b_2(b_2+\eta)=b_1b_3,\quad & b_2(b_2+\eta)=a_1c_2,\quad & b_2(b_2+\eta)=c_1a_2,\quad \nonumber\\
c_1^2=b_1d,\quad & c_2(c_2+\eta')=b_3 d,& \quad \nonumber\\
 b_1a_2=b_2a_1,\quad & c_1b_2=c_2b_1,\quad & a_2(b_2+\eta)=b_3a_1,\nonumber\quad\\
 c_1b_3=c_2(b_2+\eta) \quad &  c_1a_1=b_1(b_2+\eta),\quad & b_2b_3=c_2a_2,\nonumber\quad\\
c_1c_2=b_2d,\quad & c_1(b_2+\eta)=a_1d,\quad & c_2(b_2+\eta)=a_2d \label{reladP2def},
\eeqs
where $\eta=b^2 c \tilde \eta$ and $\eta'=b^2 u'\tilde \eta$.
The ambiguity parameters go to infinity as $\epsilon^{-1}$, so they are 
subdominant with respect to the loop variables.

We again need to shift some variables to make contact with (\ref{fstordP2}). An appropriate shift is
\beq
b_2 \rightarrow b_2 -\frac{1}{2}\eta, \qquad
c_2 \rightarrow c_2 -\frac{1}{2}\eta'.
\eeq
Of course, when plugging this into (\ref{reladP2def}), one should consider $\eta^2=0$ and use the relations (\ref{reladP2}) when necessary. Then, identifying
\beq
\eta \equiv 2\sigma,
\eeq
one recovers exactly the set of deformed equations (\ref{fstordP2}).

\end{document}